# Anomalous dispersion and negative group velocity in a coherence-free cold atomic medium

William G.A. Brown, Russell McLean, Andrei Sidorov,

Peter Hannaford and Alexander Akulshin

*Centre for Atom Optics and Ultrafast Spectroscopy,*

*ARC Centre of Excellence for Quantum-Atom Optics*

*Swinburne University of Technology, PO Box 218 Hawthorn, Melbourne, Australia*

We have observed the propagation of an approximately 35 ns long light pulse with a negative group velocity through a laser-cooled $^{85}$Rb atomic medium. The anomalous dispersion results from linear atom-light interaction, and is unrelated to long-lived ground state coherences often associated with fast light in atomic media. The observed negative group velocity ($-c/360$) in the Rb magneto-optical trap for a pulse attenuated by less than 50% is in good agreement with the value of dispersion measured independently by an RF heterodyne method. The spectral region of anomalous dispersion is between 15 and 40 MHz, which is an order of magnitude wider than that typically associated with ground-state coherences.

# Introduction

In the last decade the propagation of light through media with steep dispersion has received much attention [1]. Possible applications, particularly for slow light, are in optical telecommunications and quantum information processing [2]. As well, there is renewed interest in superluminal light propagation.

If the dispersion of a medium is anomalous $(dn/d\nu < 0)$ then the group velocity $V_g = c/[n + \nu(dn/d\nu)]$ of a light pulse can exceed the speed of light in a vacuum $c$, or even be negative if the dispersion is sufficiently steep $(\nu|dn/d\nu| \gg 1)$. The propagation time $T$ through a length $L$ of the medium, defined as $T=L/V_g$, is less than through the same length of vacuum: $\Delta T = L/V_g - L/c < 0$. This is the phenomenon of superluminal light propagation or 'fast light.' It has been long understood that information cannot be transmitted faster than $c$ [3], but it is also well established that it is possible for the peak of a smooth pulse to propagate faster than $c$ in a medium with anomalous dispersion.

As discussed in [4], the dispersion at atomic resonance depends on the linewidth $\gamma$ and the atomic density $N$, and may be expressed as a function of the linear absorption coefficient $\alpha = 2k_0\kappa$ (where $k_0$ is the vacuum wave number and $\kappa$ is the imaginary part of the refractive index):

$$\left.\frac{dn}{d\nu}\right|_{\nu_0} = \frac{\alpha(\nu_0)}{k_0\gamma}. \qquad (1)$$

Even a modest absorption resonance can have very steep associated dispersion if it is sufficiently narrow. This is one reason that following the first observations of superluminal propagation in a solid state medium [5], fast light conditions have most often been achieved in atomic media in which very narrow resonances result from ground-state coherences. These experiments exploit the steep anomalous dispersion associated with either electromagnetically induced absorption [6,7,8] or gain doublets associated with Raman scattering processes [9,10]. Fast light has also been observed via other mechanisms, including the resonant linear absorption by molecules of millimeter waves [11], coherent population oscillations in a crystal lattice [12], and the linear response of Rb atoms in a warm vapour [13], where the observed negative delay was associated with high attenuation.

Many publications have appeared in the field of superluminal light propagation in recent years. These include the direct observation of optical precursors in a region of anomalous dispersion [14] and the almost simultaneous demonstration by three groups [8,13,15] that the leading edge of the transmitted pulse never precedes that of the incident pulse, confirming that information cannot propagate with a velocity exceeding $c$.

Despite the apparent phenomenological simplicity of fast light in atomic media, some authors have offered alternative interpretations of fast light experiments. Aleksandrov and Zapasskii [16] have questioned the interpretation of experimental observations and suggested that other effects may mimic fast and slow light. Payne and Deng [17] have suggested that superluminal propagation can be explained as due not to destructive interference of different spectral components of the light pulse, but to distortion from differential gain experienced by each side of the pulse that give the impression of superluminal propagation. Macke et al [18] have formulated a set of requirements for experimental results to be considered convincing.

Motivated by such interest, we aim to explore superluminal light in simple atomic systems, where the interpretation of results should be straightforward. In this paper we observe superluminal pulse propagation using the anomalous dispersion associated with an atomic absorption resonance unrelated to long-lived atomic coherence. Macke et al [18] demonstrated from the causality principle that a large negative fractional delay in an arbitrary fast-light system requires a high-contrast absorption or gain resonance. A dense laser-cooled atomic sample with a strong absorption line meets this requirement. Using atoms in a cold atomic vapour means we are able to achieve sufficiently high anomalous dispersion without the need for ground-state atomic coherences. The connection between absorption and dispersion is well established for such a linear system. Manipulating the magneto-optical trap (MOT) in which the atoms are confined allows some control over absorption length and attenuation.

We note that superluminal pulse propagation has previously been demonstrated in a cold atomic sample [19], where steep dispersion was generated using EIT.

In the present work, the dispersive properties of the $^{85}$Rb atoms in the MOT are studied using an RF heterodyne technique [6, 20]. This was chosen because the alternative technique of using a Mach-Zehnder interferometer is very sensitive to acoustic noise and demands high stability of the optical arrangement. The dispersive medium induces a phase shift of the probe wave given by $\Delta\Phi=2\pi v L(n-1)/c$, where $L$ is the length of the atomic sample. The dependence of the phase shift on the refractive index $n$ allows us to estimate the refractive index variation $\Delta n$ and the dispersion $dn/dv$ of the medium.

## Experimental details

The optical scheme of the experiment is shown in Fig. 1. A standard retro-reflected MOT was realized in a stainless-steel vacuum chamber using a home-made laser system consisting of an extended-cavity diode laser and tapered amplifier and a second extended-cavity diode laser used as the repumping laser. The diameter of the trapping laser beams was approximately 20 mm. The diameter of the cold Rb atom cloud was approximately 3 mm.

An extended-cavity diode laser with sub-MHz linewidth tuned to the D2 absorption lines was used as a probe laser. Doppler-free saturated absorption resonances obtained in auxiliary cells not shown in Fig. 1 were used as frequency references.

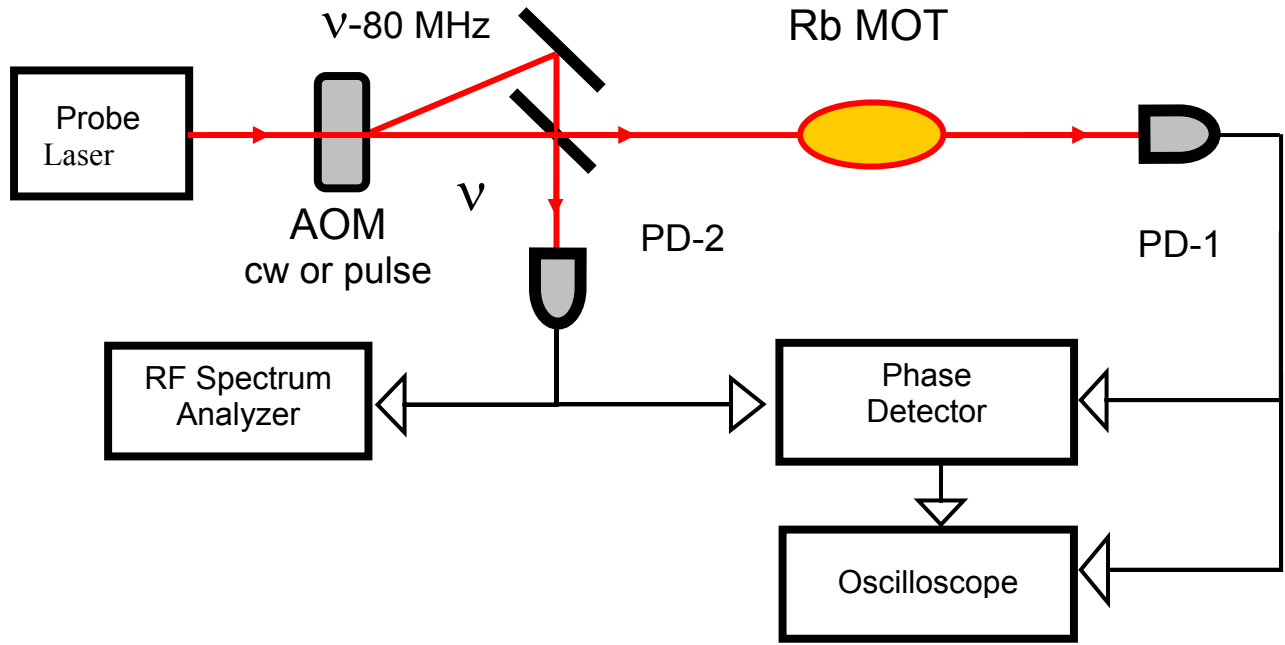

Figure 1: Scheme of the experiment.

The cw frequency-downshifted diffracted output from the acousto-optic modulator (AOM) was used as the optical signal component for the RF heterodyne scheme, while the zero-order beam was used as an off-resonant optical reference. The signal component was about four times weaker than the reference component. This was a compromise between minimizing optical pumping and light-pressure effects and maintaining a satisfactory signal to noise ratio. The intensity of the probe signal component was controlled by the rf power applied to the AOM and by neutral density filters. The maximum intensity of the linearly polarized bi-chromatic probe beam in the MOT was less than 0.1 mW/cm$^2$. The signal and reference beams were combined on a beam splitter and sent to the fast photodiodes. The AOM was also used to generate the bell-shaped optical pulses with a length of approximately 36 ns.

The probe laser propagated through the MOT at an angle of a few degrees to one of the trapping standing waves. We were able to observe the spectral dependences of absorption and refractive index simultaneously. The signal from photodiode PD-1 was sent directly to the digital oscilloscope to monitor the transmitted intensity of the probe light passing through the Rb cloud, and to the phase detector, whose output was also sent to the oscilloscope. The output of photodiode PD-2, which senses the signal and reference optical components before passing through the MOT, provides an RF reference for the phase detector.

To measure absorption and dispersion in the MOT the probe frequency was scanned across the $5S_{1/2}$ (F = 3) - $5P_{3/2}$ (F′ = 2,3,4) transitions while the trapping and repumping light were present. The atomic density in the MOT could be controlled by changing the power of the trapping radiation. The transmission of the weak probe beam at the $5S_{1/2}$ (F = 3) - $5P_{3/2}$ (F′ = 4) transition resonance was typically between 20 and 50%. Assuming that the cloud diameter is approximately the optical length of the medium $L$ and that the transmission of the weak probe obeys Beer's law $T = \exp(-\alpha_0 L)$, we can use relation (1) estimate the dispersion to be in the range - (0.9 to 2.2) × $10^{-12}$ Hz$^{-1}$, corresponding to negative group velocities $V_g$ between $-c/346$ and $-c/850$.

Figure 2 shows the transmission profiles of the bi-chromatic probe radiation through the MOT in the vicinity of the $5S_{1/2}$ (F = 3) - $5P_{3/2}$ (F′ = 2,3,4) transitions, and the output of the phase detector superimposed on a phase calibration curve. The observed phase shift is mainly due to the frequency dependence of the refractive index of the atomic cloud.

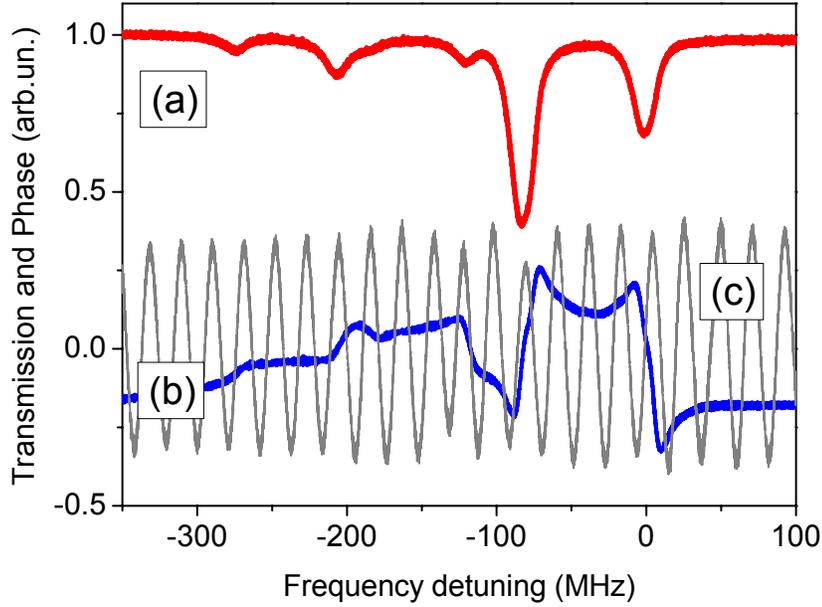

Figure 2.
(a) Typical spectral dependence of transmission of the two-component probe light through the trapped $^{85}$Rb in the vicinity of the transitions $5S_{1/2}(F = 3)-5P_{3/2}(F'=2,3,4)$.
(b) Phase detector output recorded simultaneously with the transmission profile.
(c) Phase calibration curve obtained with an auxiliary rf generator while the frequency of the probe light was scanned through the same spectral region.

Figure 2 was taken with a frequency offset of 80 MHz between the probe components, so that a single steeply dispersive spectral region produces two phase resonances separated by that offset, but with opposite polarity, as the two-component probe is scanned across the region. Because the offset is comparable with the hyperfine splitting of the $5P_{3/2}$ upper level, the observed spectrum becomes somewhat complicated as signals from different hyperfine transitions overlap. However, when the signal component was tuned to the strongest cycling transition $5S_{1/2}(F = 3) - 5P_{3/2}(F' = 4)$ the phase variation has an undistorted dispersive shape with a constant negative slope over a spectral region of 30 MHz.

The phase shift was calibrated by replacing the signal from the reference photodiode (PD-2) with the output from an auxiliary 80 MHz generator, with the amplitudes of the RF inputs to the phase detector being adjusted using an RF spectrum analyser. In this case the output from the phase detector is proportional to a harmonic function of the differential phase of the two RF inputs. A small frequency difference and random phase offset between the AOM driver and the auxiliary generator result in oscillations in the output of the phase detector. The peak-to-peak voltage of the phase detector output corresponds to a phase variation of $\pi$ and can be used to calibrate the phase signal.

## Results and discussion

Typical spectra of the probe transmission and refractive index variation in the Rb MOT are shown in Fig. 3. The experimental profiles were taken with a 220 MHz offset between the signal and reference components of the probe beam, to overcome the problem of depopulation of the MOT by the stronger reference component affecting the signal beam absorption. This was achieved by increasing the AOM drive frequency and using a double pass arrangement. A small scanning range also helped to avoid unwanted resonant interaction of the reference component with the trapped Rb atoms. From these curves we estimated values of the dispersion and atomic density in the cloud. A maximum anomalous dispersion of $dn/d\nu \approx 1.3 \times 10^{-12}$ Hz$^{-1}$ was achieved with an atomic density of $0.6 \times 10^{10}$ cm$^{-3}$ in the trap.

There is some discrepancy between the calculated profiles (Fig. 4) based on a simple two-level model applied to the three transitions and experimental observations. In the model, assuming a Lorentzian atomic response, the atomic density and power-broadened linewidth are adjusted to match the $5S_{1/2}$ (F = 3) – $5P_{3/2}$ (F′ = 4) component of the transmission profile. The refractive index is then calculated based on that atomic density and linewidth. The transmission dips on the open transitions $5S_{1/2}(F = 3) – 5P_{3/2}(F' = 2,3)$ are not as strong as they should be relative to the dip on the cycling transition $5S_{1/2}(F = 3) – 5P_{3/2}(F' = 4)$. The most likely explanation is that the ratio for the experimentally observed absorption lines is affected by hyperfine optical pumping. As well, the shape of the experimentally observed transmission profiles is somewhat different from the expected Lorentzian profile. The top of the absorption line is not as sharp, while the wings are smaller, resulting in a less steep experimental refractive index profile.

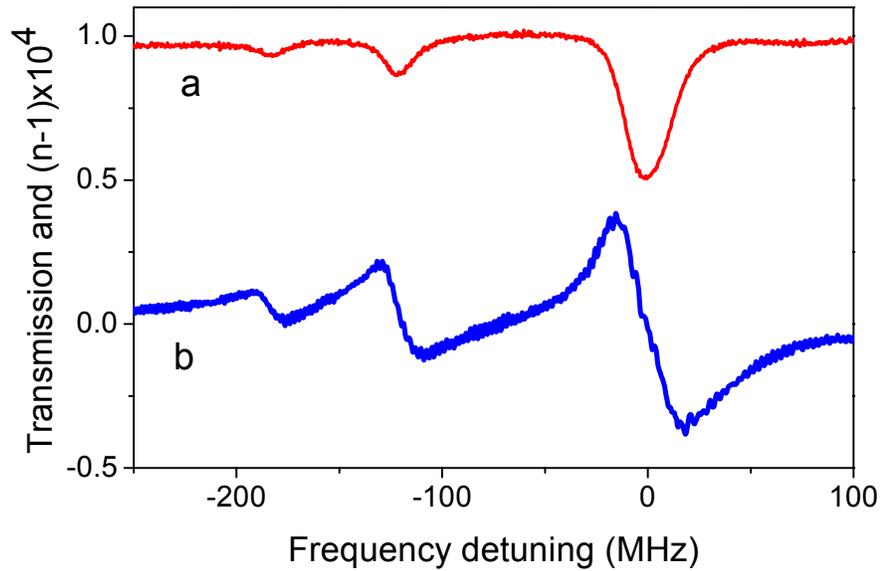

Figure 3:
Experimentally observed spectral dependences of (a) the signal component transmission through the cloud of $^{85}$Rb atoms in the MOT and (b) the refractive index of the cloud on the optical transitions $5S_{1/2}(F = 3)-5P_{3/2}(F'=2, 3, 4)$.

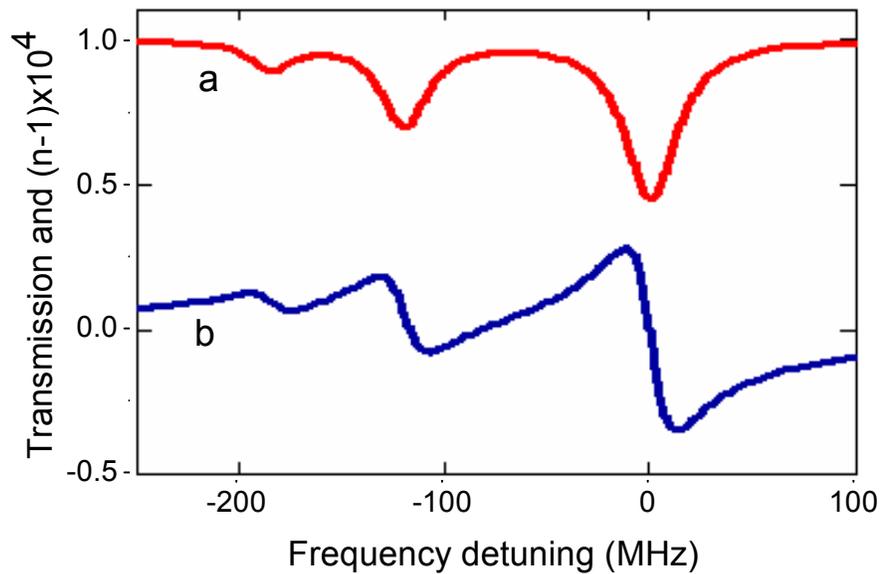

Figure 4:
Calculated transmission (a) and refractive index (b) profiles of trapped $^{85}$Rb atoms on the optical transitions $5S_{1/2}(F = 3)-5P_{3/2}(F'=2, 3, 4)$.

To directly observe pulse propagation through the Rb MOT the AOM was used in a pulsed mode. The frequency of the diffracted component of the probe laser was tuned to the transmission minimum on the $5S_{1/2}(F = 3) - 5P_{3/2}(F' = 4)$ transition, where the dispersion is negative and constant. The optical pulse of duration $\Delta t \approx 36$ ns has a spectral width $1/2\pi\Delta t \approx 4.5$ MHz, which is less than the region of negative dispersion.

Figure 5 shows a pulse of off-resonant light and a (normalized) pulse of resonant light, both detected after transmission through the Rb MOT by the photodiode PD-1 and recorded after averaging on the oscilloscope. The attenuation of the resonant pulse by the atomic sample was only about 50%. The leading edge and the top of the resonant pulse are shifted by up to $\Delta T \approx 4.5$ ns relative to the reference, while the trailing edge is advanced by 2.8 ns. The advancement estimated from Gaussian fitting is also $\Delta T \approx 2.8$ ns. Assuming a cloud size of about 3 mm yields a group velocity for the pulse of $V_g \approx - 0.83 \times 10^6$ ms$^{-1} \approx -c/360$. This is in good agreement with the measured dispersion of $dn/d\nu \approx 0.9 \times 10^{-12}$ Hz$^{-1}$. The advance decreases in a more dilute Rb MOT as expected. The average value for the advance, $\Delta T \approx 3.6$ ns, represents a significant fractional advance of $\Delta T/\Delta t \approx 10\%$. Larger fractional advances should be achievable in a more dense compressed MOT, but with stronger pulse attenuation and degraded signal to noise.

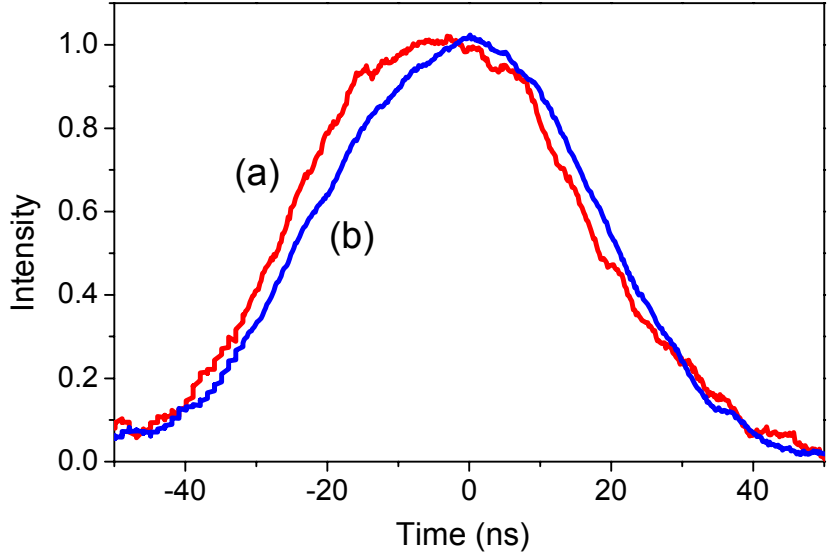

Figure 5:
Normalized resonant (a) and off-resonant (b) pulses after propagation through the $^{85}$Rb MOT. The frequency of the linearly polarized resonant probe was tuned on the transitions $5S_{1/2}(F = 3) - 5P_{3/2}(F' = 4)$. The attenuation of the resonant pulse after transmission is less than 50%.

The result is consistent with the common observation [8,9,10,13] that it is harder to achieve the same fractional advance of a light pulse as fractional delay, particularly with low distortion. While delays of several times the pulse width are common, Macke *et al.* [18] have estimated the largest advance that could be attained in a realistic experiment, and claim that in any experimental system advances exceeding two times the full width at half maximum of the pulse intensity profile are unattainable.

The observed advance is in reasonable agreement with the dispersion of $dn/d\nu \approx -1.3 \times 10^{-12}$ Hz$^{-1}$ estimated from the RF heterodyne measurements.

## Conclusion

We have observed superluminal pulse propagation for a 36 ns optical pulse propagating through cold $^{85}$Rb atoms in a magneto-optical trap, with a pulse advancement of about 3.6 ns for a pulse attenuated by about 50%. The anomalous dispersion in a spectral region up to 40 MHz that gives rise to the superluminal propagation is associated with the linear atom-light interaction of a simple absorption resonance, with no long-lived atomic coherence present. The observed negative group velocity of $-c/360$ is in good agreement with the values of anomalous dispersion measured by an RF heterodyne technique and estimated from a two-level model applied to three two-level transitions. This suggests that the concept of group velocity is meaningful for linearly-responding fast-light media and basic features of pulse propagation with negative group velocity can be predicted based on a linear two-level model.